\documentclass[runningheads]{cl2emult}

\usepackage{makeidx}  
\usepackage{graphicx} 
\usepackage{subeqnar} 
\usepackage{multicol} 
\usepackage{cropmark} 
\usepackage{physbb}   

\usepackage{sidecap_rudy}
\usepackage{psfig}

\makeindex            

\begin{document}
\title*{Relativistic Jets from X-ray binaries
\footnote{To be published in `ASTROPHYSICS AND COSMOLOGY : a
collection of critical thoughts', Springer Lecture Notes in Physics}
}
\toctitle{Relativistic Jets from X-ray binaries
}
%
%
\titlerunning{Relativistic Jets from X-ray binaries
}
%
\author{R.~P.~Fender}
\authorrunning{R.~P.~Fender}
%
%
\institute{
		Astronomical Institute `Anton Pannekoek'\\
		and Center for High-Energy Astrophysics,\\
		University of Amsterdam,
		Kruislaan 403,\\
		1098 SJ Amsterdam,
		The Netherlands
}

\maketitle              

\begin{abstract}

In this review I summarise the status of observational research into
relativistic jets from X-ray binaries, highlighting four areas in
particular: (i) How relativistic are the jets ?, (ii) The disc : jet
coupling, (iii) the nature of the underlying flat spectral component,
and (iv) the relation between jets from black holes and those from
neutron stars. I have attempted to discuss the extent of our (limited)
physical understanding, and to point the way towards relevant new
observational tests of the various phenomena.

\end{abstract}

\section{Introduction}

At the time of writing, some 200 -- 250 X-ray binaries, systems in
which a neutron star or black hole is accreting material from a
companion star, are known. Approximately 20\% of these sources have
been detected at radio wavelengths, and in several cases high
resolution radio observations have resolved this emission into
jet-like structures, sometimes with components moving at relativistic
velocities.  All of these relativistic jets emit primarily via
incoherent synchrotron emission from very high energy electrons
spiralling in magnetic fields (although other emission mechanisms may
contribute to the weak, flat spectral components which are sometimes
observed).  Detailed reviews of the synchrotron emission process and
radio emission from X-ray binaries in general can be found in
\cite{HM91,HH95}.

\begin{figure}
\begin{center}
\includegraphics[width=.7\textwidth,angle=270]{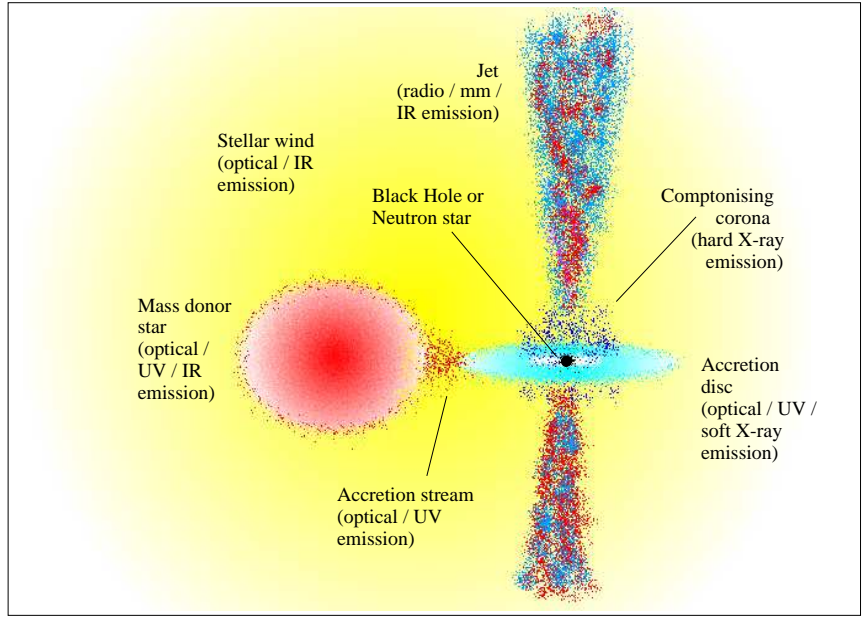}
\end{center}
\caption[]{
A sketch of the various emitting regions in an X-ray binary, as we
currently envisage them. The jet is believed to propagate from the
inner accretion disc, more or less perpendicular to the binary plane.
}
\label{ps2}
\end{figure}

\begin{SCfigure}
\includegraphics[width=.55\textwidth]{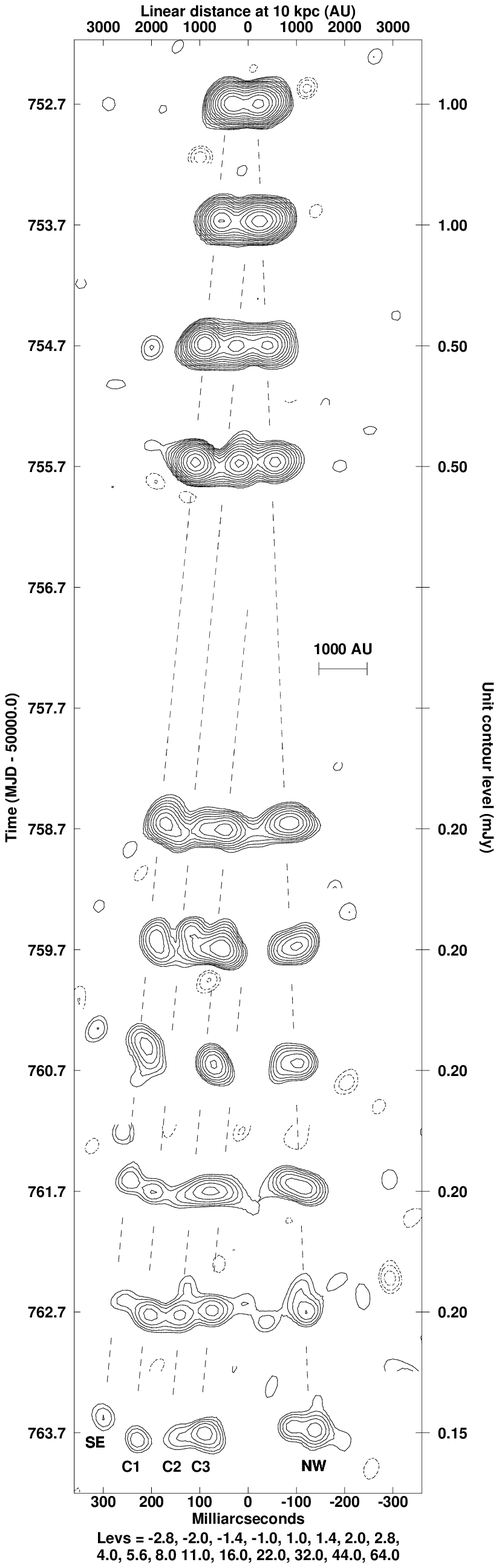}
\caption[]{ A sequence of ten epochs of radio imaging of relativistic
ejections from the black hole candidate X-ray binary GRS 1915+105
using MERLIN at 5 GHz.
The figure has been rotated by 52 degrees to form the
montage. Contour levels increase in factors of $\sqrt{2}$ from the
unit contour level indicated at the right hand side of each image.
Components SE, C1, C2 \& C3 are approaching with a mean proper motion
of $23.6 \pm 0.5$ mas d$^{-1}$. Component NW is receding with a mean
proper motion of $10.0 \pm 0.5$ mas d$^{-1}$ and corresponds to the
same ejection event which produced approaching component SE. For an
estimated distance to the source of 11 kpc the approaching components
have an apparent transverse velocity of $1.5c$. Assuming an intrinsically
symmetric ejection and the standard model for apparent superluminal
motions, we derive an intrinsic bulk velocity for the ejecta of
$0.98^{+0.02}_{-0.05}c$ at an angle to the line of sight of $66 \pm 2$
degrees (at 11 kpc).  The ejections occurred after a 20-day `plateau'
during which the X-ray emission was hard and stable and the radio had
an inverted spectrum.  The first two ejections were punctuated by four
days of rapid radio oscillations, indicative of an unstable inner
accretion disc being repeatedly ejected \cite{FPBN97,PF97,M98,E98,FP98}. 
The apparent curvature of the jet is probably real, although the cause
of the bending is uncertain.
A detailed presentation and
discussion of these results may be found in \cite{Fen99}.  }
\label{ps1}
\end{SCfigure}

In this review I will not spend time discussing sources individually,
but rather try to concentrate on areas in which the physics of what is
occurring in general might be probed. This is due in part to detailed
discussion of individual sources in previous reviews
(e.g. \cite{HH95,FBBW97,ziol97,MR99}) and in
part to a feeling that the jet mechanism is to a large degree
independent of what is occurring beyond the inner parts of the
accretion disc (leading to a disparate collection of binary systems
with common jet characteristics).

I also have to stress that much of the following discussion has been
heavily based upon discussions on the nature of AGN jets which have
been ongoing for more than 20 years, and that many of the problems I
am attempting to address were first recognised by the AGN community
more than a decade ago.

Fig. 1 attempts to summarise our current ideas regarding where the
different forms of radiation from an X-ray binary physically
originate. The jets are believed to form close to the accreting
compact object and to propagate in opposite directions away from the
compact object along the symmetry axis of the inner accretion disc
(although it may well be instead the rotation axis of the black hole,
which in the case of GRO J1655-40 {\em may} be misaligned with the
orbital axis by $\sim 15^{\circ}$ \cite{ob97}).
Fig. 2 is an example of {\em real} radio images of jets from which,
combined with optical, infrared and X-ray flux monitoring (e.g. Figs 4
\& 5), and a smattering of theory, we draw such inferences.

\section{How relativistic ?}

An important and obvious point to raise concerning the jets from X-ray
binaries is {\em how} relativistic are the bulk motions ? This is a
fundamental question for the theory and energetics of these systems,
and is not as well constrained by observations as might be imagined
(only SS 433, with $\beta = v /c = 0.26$ has a confidently determined
velocity).  For the MERLIN observations of GRS 1915+105 (Fig 2;
\cite{Fen99}) the derived bulk Lorentz factor $\gamma =
(1-\beta^2)^{-1/2}$, under the assumption of an intrinsically
symmetric ejection, is 1.8 at 9 kpc (a likely lower distance to the
source), 2.6 at 10 kpc, and asymptotically approaches infinity at
$11.2 \pm 0.8$ kpc (the derived upper limit to the distance). This
large uncertainty in $\gamma$, and its sensitivity to assumed
distance, is illustrated in Fig 3.  It seems unlikely that we are
going to be able to measure the distance of the system accurately
enough to place an upper limit on the Lorentz factor, so what other
methods can we use ?

\begin{figure}
\begin{center}
\includegraphics[width=.6\textwidth,angle=270]{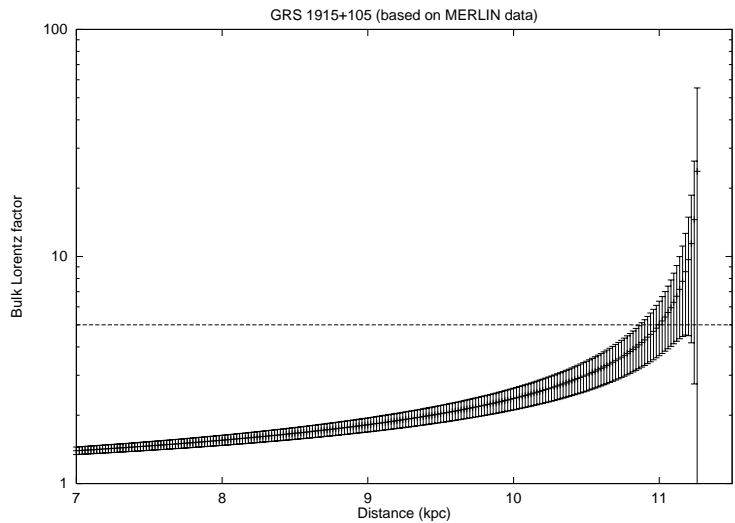}
\end{center}
\caption[]{
The variation in derived bulk Lorentz factor for GRS 1915+105 based
upon the MERLIN observations of \cite{Fen99}. The horizontal line
corresponds to $\gamma=5$ (the solution for a distance of 11 kpc).
}
\label{ps2}
\end{figure}

In \cite{Fen99} a lower limit to the energy budget of the ejecta
observed from GRS 1915+105 with MERLIN was calculated (based upon
several assumptions which are outlined in the paper), and it was shown
that, at a distance of 11 kpc, the minimum power requirement for
generation of internal energy, assuming the primary ejection was
generated in $\sim 12$ hr, was $4 \times 10^{38}$ erg s$^{-1}$. This
power requirement increases by a factor of four when multiplied by
$(\gamma-1)$ to account for the kinetic energy, and by a further
factor of two if there is a cold proton for each electron. Thus the
total power requirement is of order the Eddington luminosity for a $10
M_{\odot}$ black hole.

Based upon the equations presented in \cite{Fen99} it can be shown
that the total power (internal + kinetic), $P \propto \gamma^r$, where
$r \sim 3.5$ (for discrete ejections and a spectral index $\sim
-0.8$).  The dependence of the total mass of protons on $\gamma$ is
weak and so we do not consider it here.  Assuming that the
black hole in GRS 1915+105 cannot generate more than
$10^{42}$ erg s$^{-1}$, we can derive a maximum Lorentz factor of
$\sim 30$. A more conservative limit on the power of $10^{41}$ erg
s$^{-1}$ leads to a maximum Lorentz factor of $\sim 15$.

So we find that the bulk Lorentz factor for GRS 1915+105 alone can
only be constrained with any confidence between 1.4 ($\beta=0.7$) and
30 ($\beta=0.9995$ !), with even these numbers based upon several
assumptions.  For the other highly relativistic jet sources (in
particular GRO J1655-40, but also probably XTE J1748-288 and Cyg X-3)
the data are aguably even less constraining than they are for GRS
1915+105. In addition, only for SS 433 is there strong evidence for
many independent ejections having the same bulk velocity, although it
is encouraging that VLBA observations of GRS 1915+105 \cite{dhaw98}
at a different epoch to our MERLIN observations
\cite{Fen99} derive the same proper motion for the approaching
component, considerably higher than the earlier VLA observations
(which may have been hampered by insufficient angular resolution).
This range in Lorentz factor is similar to the range inferred for AGN
(e.g. \cite{vc94}) from VLBI imaging, although lower than the very
high values of $\gamma$ (up to 100) apparently required to explain
some AGN rapid variability (e.g. \cite{wag97}). Note that under the
assumption that $\gamma \sim 2.5$ for GRS 1915+105 and GRO J1655-40
(i.e. before the more recent MERLIN and VLBA results), models have
already been developed to explain why jets from X-ray binaries have a
lower terminal velocity than those from AGN \cite{ll96,rh98}.
It would clearly be of great importance to determine the distance of a
superluminal source to be significantly less than the maximum allowed
(for $v=c$) and thereby place an upper limit on $\gamma$.

\section{The disc : jet coupling}

One of the most active areas of study over the past couple of years
has been the relation between the accretion disc (the X-ray source)
and the jet (the radio source). In many, maybe all, systems, some
dramatic change in the jet (whether inferred from radio emission or
directly imaged), is found to have a corresponding event in
X-rays. This link between X-ray state changes and radio emission was
noted already by \cite{HH95}.

On the most general level, it is found that most bright X-ray
transients, including the classical soft X-ray transients (SXTs) such
as A0620-00, are accompanied during X-ray outburst by a radio outburst
(see e.g. \cite{HH95,Kuul99}). Since these outbursts are
believed to result from rapid increases in mass accretion rate onto
the compact object as a result of a disc instability, it is clear in
this instance that a significant change in the accretion disc
structure and mass accretion rate are coupled to the production of a
radio event. It seems very likely that such events correspond to
ejections of plasmons, possibly at relativistic velocities
\cite{Kuul99}, although an outflow from a `classical SXT' (if such a
thing really exists) has yet to be unambiguously resolved.

\begin{figure}
\begin{center}
\includegraphics[width=.95\textwidth,angle=270]{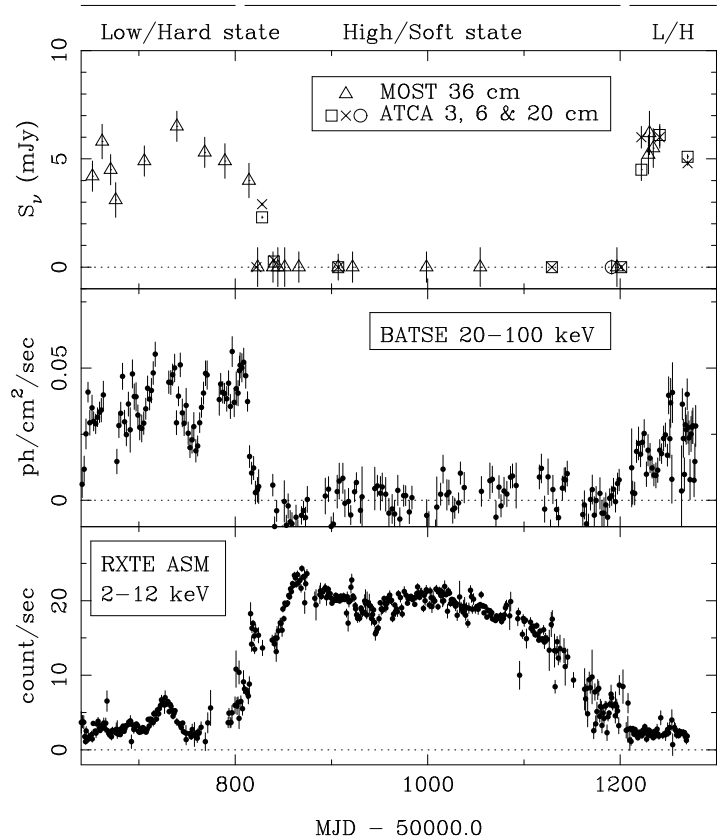}
\end{center}
\caption[]{
Radio, hard- and soft-X-ray monitoring of GX 339-4 before,
during and after the 1998 high/soft state. The radio emission is
reduced by more than a factor of 25 during this state. Unsually
optically thin emission ($S_{\nu} \propto \nu^{-0.4}$)
is observed just before and after the state
changes, probably corresponding to discrete ejection events.
From \cite{Fen99b}.
}
\label{ps1}
\end{figure}

\begin{figure}
\begin{center}
\includegraphics[width=.65\textwidth,angle=90]{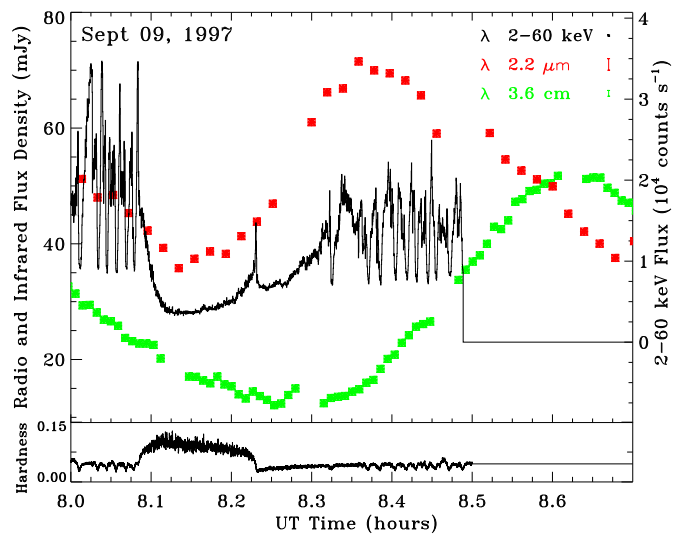}
\end{center}
\caption[]{
Simultaneous X-ray, infrared and radio observations of GRS
1915+105. Following a dip in the X-rays (around UT = 8.1 hr), 
modelled as the disappearance
of the inner 100 km or so of the accretion disc, the infrared rises
and is followed shortly thereafter by a rise in the radio emission
with the same pulse shape. This is interpreted as ejection of some
fraction of the inner accretion disc and subsequent observation of
synchrotron emission from cm -- $\mu$m wavelengths from the new
plasmon. Figure from \cite{M98}; see also \cite{FPBN97,PF97,E98,FP98}.}
\label{ps1}
\end{figure}

In the case of the more persistent BHC systems, Cyg X-1 and GX 339-4,
amongst the best candidates for quasi-continuous weak jets (see
below), two effects indicate a strong disc : jet coupling. Firstly,
the X-ray and radio emission are correlated on long ($\geq$ days)
timescales during the most common low/hard X-ray state
\cite{Hann98,PFB99}, including during brief periods of flare-like
activity in Cyg X-1 \cite{PFB99,brock99}. Secondly and more
dramatically, the radio emission abruptly drops below detectable
levels (by a factor of $\geq 25$ in the case of GX 339-4) during
transitions to high(er)/soft(er) states (Fig 4; \cite{BM76,Fen99b}).
Such state transitions are believed to physically correspond to major
changes in the geometry and relative importance of the inner accretion
disc and Comptonising corona, and it is clear that the low/hard state
generates a flat-spectrum, persistent radio source, which is likely to
be a conical quasi-steady jet, whereas the high(er)/soft(er) states do
not \cite{Fen99b}. Whether the high(er)/soft(er) states simply do not
generate an outflow, or whether the radio emission is quenched in some
way (e.g. via Compton cooling by a much increased density of soft
X-ray photons) is unclear at present.

Perhaps the most dramatic example of disc : jet coupling has been in
GRS 1915+105, in which radio -- infrared oscillations with periods of
10 -- 60 min have been found to follow X-ray dips (see Fig 5). These
radio -- infrared oscillations have been interpreted as synchrotron
emission from repeated small ejections suffering strong adiabatic
expansion losses \cite{FPBN97,PF97,E98,M98,FP98}, while the X-ray dips
have been interpreted as the repeated disappearance of the inner
($\sim$ few 100 km of the) accretion disc \cite{Bell97}. The obvious
conclusion to be drawn is that some of the inner disc is being
accelerated and ejected away from the system. The fraction of the
accretion rate which is ejected has been estimated at roughly 10\%
\cite{FP98} but the uncertainties in this calculation, due to
incomplete knowledge of the spectrum, size, filling factor of the
ejecta and the comparative reality of numbers derived from X-ray
spectral fits, etc., are very large. Still, it is of great interest to
know whether or not black holes remove the majority of matter that
they attract from the Universe, or whether they spit it back at
relativistic velocities!

In addition to a soft X-ray (disc) : radio (jet) coupling, there is
also clearly a related hard-X-ray -- radio coupling (e.g. Fig 4;
\cite{Fen99b} and references therein). In particular broad-band hard
X-ray emission (i.e. 20--100 keV) as observed with GRO/BATSE is very
well correlated with radio emission in many cases.  It is currently
believed that the hard X-ray emission is a result of inverse Compton
scattering of lower-energy photons by a corona of high-energy
electrons (e.g. \cite{pou98}). Given that both the jet and the corona
are inferred to arise in the vicinity of the inner accretion disc, the
{\em inferred} composition of the corona as being a population of
high-energy electrons, and the {\em observed} transport of
relativistic electrons by the jet (we estimated a mean Lorentz factor
for the synchrotron-emitting electrons of 240 for the ejections from
GRS 1915+105 observed with MERLIN), it seems that the corona and jet
must be inextricably linked. In both \cite{pou98} and \cite{Fen99b} it
is suggested that the corona is simply the base of the jet; however
Comptonisation modellers have yet to take this suggestion seriously
and incorporate it into the geometries and bulk dynamics of their
models. While it seems likely that the high-energy electrons
responsible for the synchrotron emission are simply the high-energy
(nonthermal) tail of the same energetic population of electrons
responsible for the Comptonisation, more detailed calculations of both
populations are required before we can be certain.

\section{Compact, flat-spectrum cores}

For a simple, single synchrotron source we expect to observe a spectral
index ($\alpha$, where flux density $S_{\nu} \propto \nu^{\alpha}$) of
+2.5 below frequencies at which the source is self-absorbed, and in
the range $-0.5$ to $-1$ at frequencies at which the source is optically
thin (see e.g. \cite{HM91} for more details). However, in many cases
of persistent or repetitive radio emission from X-ray binaries, there
appears to be an underlying flat-spectrum ($\alpha=0$) component.
During outbursts the emission is dominated by far brighter components
which rapidly evolve to an optically thin state (presumably as they
expand). For example, during relative quiescence Cyg X-3 shows a flat
spectrum between cm -- mm wavelengths \cite{Fen95,Og98}, with
little evidence for either high- or low-frequency cut-offs. Even more
extreme, during periods of radio oscillations GRS 1915+105 appears to
show a flat spectrum from cm -- $\mu$m wavelengths
\cite{FPBN97,M98,FP98}.

\begin{figure}
\begin{center}
\includegraphics[width=.75\textwidth,angle=270]{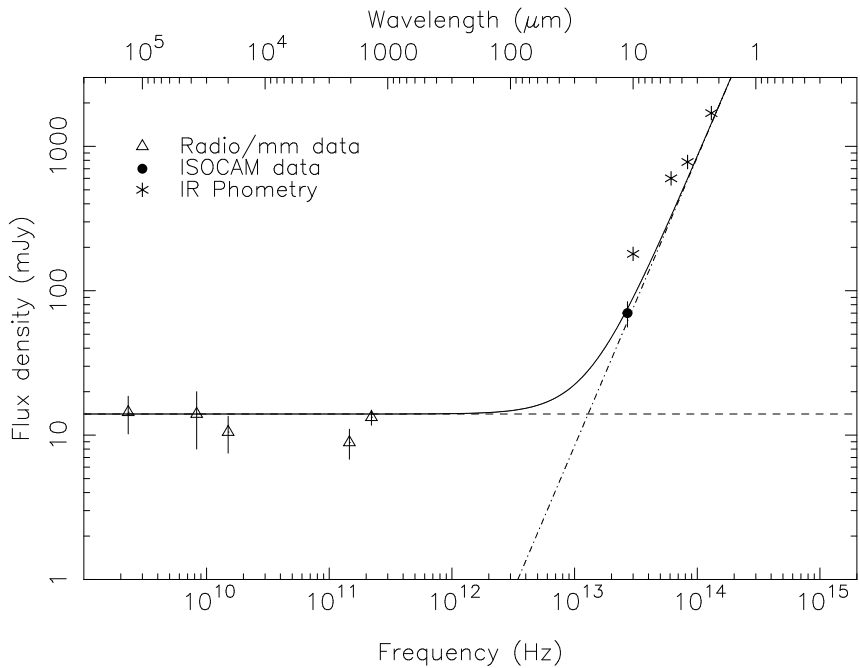}
\end{center}
\caption[]
{
Illustration of the radio--mm flat spectral component in Cyg X-1,
from \cite{Fen99c}. At
wavelengths shorter than $\sim 30 \mu$m the component is swamped by
thermal emission from the OB companion star. We have no clear
determination of either high- or low-frequency cut offs to this
spectral component, and its origin remains uncertain. Brighter,
analogous flat-spectrum components observed from Cyg X-3 and GRS
1915+105 appear to extend to at least $\sim 2 \mu$m.
}
\label{eps1}
\end{figure}

By analogy with AGN, it seems likely that the flat spectra of these
quiescent components may orginate in partially self-absorbed
synchrotron emission from conical jets, with higher frequencies probing
the smaller and brighter regions towards the base of the jet
(e.g. \cite{MG85,HJ88,falcke95,falcke96}).
Amongst the best examples of persistent flat spectrum cores which
experience little `contamination' from optically thin components, and
therefore probably of continuous jets, are the persistent black hole
candidates Cyg X-1 and GX 339-4 (\cite{PFB99,FSNT97}). In
particular, in Cyg X-1 the evidence for a compact jet seems
compelling:

\begin{itemize}
\item{A flat radio spectrum \cite{PFB99}, recently detected to extend
to mm wavelengths (Fig 6; \cite{Fen99c}).}
\item{A $\sim 20$\% modulation at the 5.6-d orbital period at
15 GHz, with minimum near superior
conjunction of the compact object, and increasing delays and
decreasing fractional modulation at lower frequencies \cite{PFB99,brock99}.}
\item{Direct imaging of structure with the VLBA \cite{Stir98,dlf99}.}
\end{itemize}

Futhermore, the recent detection of low-level ($\sim 2$\%) linear
polarisation of the radio emission from GX 339-4 \cite{Cor99}
opens up the possibility of probing the magnetic field structure
within such compact jets. Such a low level of linear polarisation was also
detected from the `second stage' optically thick radio component
associated with the X-ray transient V404 Cyg \cite{HH92}. This
component also showed `radio QPO' possibly similar to those observed
from GRS 1915+105 (see above) and may in fact have represented a
temporary transition to a compact jet following the major outburst (as
opposed to the expanding shell model proposed by \cite{HH92}).
Detection of a comparable level of linear polarisation from the
(inferred) compact jet in Cyg X-1, and its variation with the orbital
cycle, would be very interesting.  The radio properties of Cyg X-1 and
GX 339-4 are similar to those of the neutron star Z-sources, low-mass
X-ray binaries accreting at or near to the Eddington luminosity,
suggesting that these sources may also be producing quasi-stationary
outflows (see below). 

In addition the relation of the flat-spectrum radio--infrared
oscillations of GRS 1915+105 to the (apparently) more continuous
flat-spectrum jets discussed above is unclear -- possibly the
oscillations are some kind of intermediate state or pulsed jet. The
`plateau' states in GRS 1915+105 (\cite{Fen99} and references therein)
appear to be characterised by 
luminous flat spectrum radio emission which may be
a brighter (larger ?) version of the flat spectrum cores in e.g. Cyg
X-1. If this is the case GRS 1915+105 appears to produce steady jets,
pulsed jets and major (optically thin) ejections at different times.

It is important to stress that neither high- nor low-frequency cutoffs
have been found in the flat-spectrum component of {\em any} X-ray
binary. The extension of a synchrotron spectrum to high (e.g. infrared
in GRS 1915+105) frequencies dramatically increases the {\bf power}
required for the generation of high energy electrons and magnetic
field. The extension of the spectrum to low radio frequencies
increases the {\bf number} of electrons required, highly significant
for the mass-flow rate if each electron is accompanied by a proton. It
may turn out to be extremely difficult to observe such cut-offs
however: in the case of Cyg X-1 for example, thermal emission will
dominate the flat synchrotron spectrum at $\lambda \leq 30\mu$m
 (Fig 6; \cite{Fen99c}); in
the case of GRS 1915+105 extreme optical extinction will probably
preclude the discovery of the optical counterparts of the
near-infrared synchrotron oscillations; and in all cases radio
observations at frequencies below $\sim 100$ MHz are generally
difficult to make. 

It should be stressed that while the flat-spectrum component from
X-ray binaries is reminiscent of that observed from `flat-spectrum'
AGN, it is in fact {\em much flatter} \cite{Fen99c}, with a spectral
index very close to zero over several decades in frequency. So, while
it is natural to try and apply the compact synchrotron-emitting jet
models developed for AGN, it is by no means certain that they are
relevant.  At least two additional possibilities exist :

\begin{itemize}
\item{Optically thin emission from an electron spectrum which is 
much harder than that observed during major
outbursts \cite{Mel97,Wang97}. However the frequency-dependent time
delays observed in e.g. GRS 1915+105 argue against this
interpretation, as they imply significant optical depth.}
\item{(Nonthermal) free-free emission which should produce a perfectly
flat spectrum. It is conceivable that this process dominates over
synchrotron in the most compact and dense parts of the jet, but that
during discrete ejections, as the plasmons expand, synchrotron
emission (which has a weaker dependence on the electron number
density) comes to dominate.}
\end{itemize}

Important future observations will include discovery of high- or
low-frequency cutoffs in the flat spectral component, variability
timescales and measurements of polarisation.

\section{Black holes and neutron stars}

The majority of the highly relativistic Galactic jet sources
potentially (perhaps even probably) contain black holes. Furthermore,
a more thorough survey of the literature reveals that it is the BHC
X-ray transients which are most likely to be accompanied by strong
radio outbursts. While some neutron star transients have also produced
radio outbursts, e.g. Aql X-1 (at least once) and Cen X-4 \cite{HH95},
no transient which has been demonstrated unequivocally to harbour a
neutron star has ever matched the very bright ($\geq 1$ Jy) radio flux
densities recorded from BHC transients such as V404 Cyg and GRO
J1655-40.

\begin{figure}
\begin{center}
\includegraphics[width=.8\textwidth]{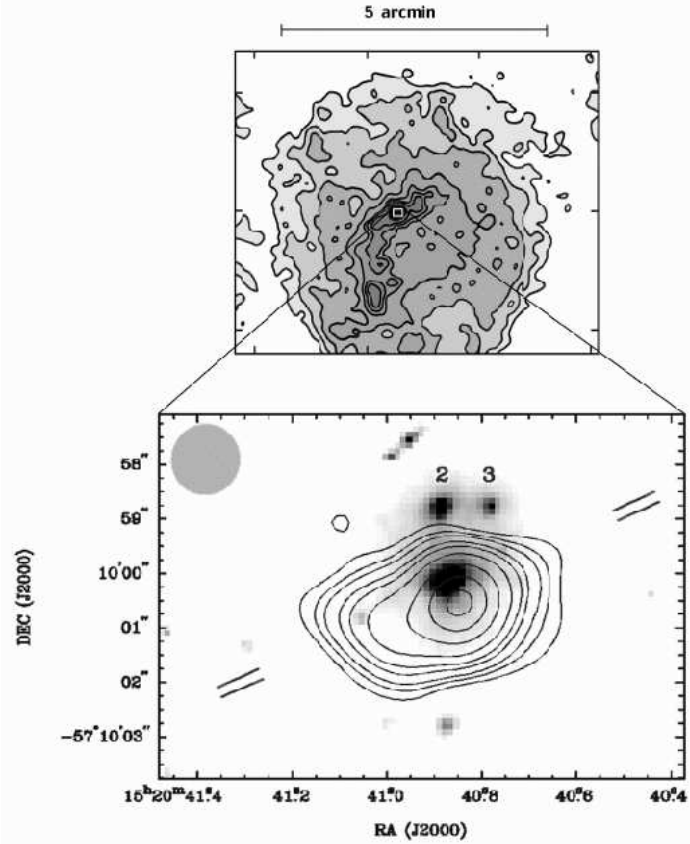}
\end{center}
\caption[]
{
An asymmetric arcsec-scale radio jet from the neutron star X-ray
binary Cir X-1. The arcsec-scale jet aligns with larger scale
collimated structures in the surrounding nebula, implying transport of
matter from the binary to regions over a parsec away. The asymmetry of
the jet, if due to Doppler boosting of an intrinsically symmetric
outflow, indicates bulk motion at $\geq 0.1c$. From \cite{Fen98}.
}
\label{eps1}
\end{figure}

The brightest radio source associated with a neutron star in an X-ray
binary is that of Cir X-1. One of the persistently brightest X-ray
sources in the sky (although with a quite dramatically changing X-ray
light curve -- \cite{tw99}), in the 1970s Cir X-1 was
detected as a $\geq 1$ Jy radio source, with outbursts periodic at the
16.6-day orbital period. However, since that time the radio source has
weakened considerably, and at present the radio core is detected at
$\leq 10$ mJy with more flux at lower frequencies arising in the
complex surrounding synchrotron nebula \cite{Stew93}. Within this
nebula collimated arcmin-scale structures are evident, aligned with
which is an arcsec-scale asymmetric jet \cite{Stew93,Fen98}
(Fig 7). Naively intepreting the asymmetry of the jet as relativistic
aberration of an intrinsically symmetric ejection (or, more likely,
multiple ejections), a velocity for the outflow of $\geq 0.1c$ was
derived. Further observations to confirm that this structure is
dynamic, and hopefully to track its expansion and hence directly
measure outflow velocities, are very important in seeking to establish
whether or not neutron stars can also produce highly relativistic
outflows. Recent VLBA observations of another neutron star source, Sco
X-1, have also revealed strong evidence for an outflow \cite{Brad97}.
It is interesting to note that several authors (e.g. \cite{liv97})
have suggested that the velocity of jets from neutron star X-ray
binaries (a class in which SS 433 and Cyg X-3 were often included
without any clear justification) would always be $\sim 0.3c$ and those
from black holes would always be $\geq 0.9c$, this dichotomy
reflecting the escape speed from near the surfaces of these two types
of objects (supported as a concept by the velocities observed from
protostars and white dwarfs). It seems that within the next couple of
years we will be in a position to directly test this assertion.

Beyond transient and unique sources, three clear classes of
persistently bright neutron star X-ray binaries exist, the Z and Atoll
sources \cite{vdk95} and X-ray pulsars \cite{wnp95}.  The Z and Atoll
sources are believed to contain low ($B \leq 10^{10}$ G) magnetic
field neutron stars, the X-ray pulsars much higher field ($B \geq
10^{11}$ G) neutron stars.  The Z sources are accreting near the
Eddington limit and include Sco X-1 and possibly also Cir X-1
(although the latter is still unique in many respects). While faint
(mostly lying at distances $\geq 8$ kpc) their radio properties appear
to be very similar to those of the BHCs GX 339-4 and Cyg X-1,
discussed above. Combined with the direct imaging evidence for
outflows from Sco X-1, this suggests that the Z sources also have weak
jets.  The Atoll sources, which have timing properties similar to the
BHCs are {\em not} in general observed as radio sources; only GX 13+1
is confirmed as a bright radio emitter (and may in fact be a hybrid
Z/Atoll source). No strong-field X-ray pulsar systems have ever been
detected as radio synchrotron sources and it seems likely that they do
not produce jets, probably as a result of disruption of the inner
accretion disc by the strong field of the neutron star \cite{Fen97}.

Future careful comparisons of the nature of radio emission from the
various populations of neutron stars with different magnetic fields,
and with the black holes, may give us fundamental clues to the puzzle
of jet formation (is a high accretion rate and a weakly magnetised
compact object all that is required ?).

\section{Conclusions}

In this review I have attempted to summarise the state of some
particular problems in observational research into relativistic
outflows from neutron stars and black holes in X-ray binary systems. I
have been able to do no more than sketch an outline of the work going
on in this exciting field and, due to space limitations, have been
forced to omit several major areas which deserve full discussions on
their own. Significantly among these are the fundamental questions of
jet structure (continuous or discrete ?), jet composition (e$^-$:e$^+$
or e$^-$:p$^+$ ?) and the physics of jet formation !

Much work remains to be done in this field, and it is this author's
(by no means original !) feeling that jets will turn out to be
ubiquitous wherever there is an (inner) accretion disc and a high
(approaching Eddington ?)  accretion rate.  The sooner that the
coupled disc : jet system is considered as a single entity and not as
two distinct problems by the high and low(er) energy communities
respectively, the better.

I have no hesitation in recommending the excellent volumes `Beams and
Jets in Astrophysics' \cite{BJ91} and `X-ray binaries' \cite{XRB95} as
reference points for much of the theory and observational data
presented in this review.

\section*{Acknowledgements}

I would like to thank the following people for many useful discussions
which have contributed in some way to the contents of this review: Guy
Pooley, St\'ephane Corbel, Kinwah Wu, Lars Bilsten, Alistair Stirling,
Eric Ford, Michiel van der Klis, Jan van Paradijs, Gijs Nelemans, Rudy
Wijnands, Peter Jonker and many others at the Astronomical Institute
`Anton Pannekoek'.

\clearpage
\addcontentsline{toc}{section}{Index}
\flushbottom
\printindex

\end{document}